\documentclass[a4paper]{jpconf}
\usepackage{graphicx}
\usepackage{amsmath}
\usepackage{amssymb}
\newcommand{\GHZ }{{\rm GHZ}}
\newcommand{\perm }{{\rm perm}}

\newcommand{\Id}{{\rm Id}}

\newcommand{\eps }{\varepsilon}

\begin{document}
\title{Multimode entanglement for fermions}

\author{M~Rouleux}

\address{Aix-Marseille Universit\'e, Universit\'e de Toulon, CNRS, CPT, Marseille, France}

\ead{rouleux@univ-tln.fr}

\begin{abstract}
We are motivated by tripartite entanglement for fermions. While GHZ or W states involve 3-fold
intrication, we consider here piecewise intrication of 3 fermions in ${\bf C}^2$, namely of type $ab+bc+ca$. 
Before interaction with Stern-Gerlach apparatus, qu-bits are distinguishable; at the output however they turn into 
un-distinguishable
particles, whose anti-symmetric wave function is of the form $\det(b-a,c-a)$
(affine determinant).
More generally,
$d+1$ intricated fermions in ${\bf C}^d$ can be represented by the anti-symmetric wave function  $\det(a_1-a_0,a_2-a_0,\cdots,a_d-a_0)$.
We investigate also properties of affine Slater determinants, as expectation values or reduced density matrices.
\end{abstract}

\section{\bf Introduction}
\smallskip
Multimode entanglement has been widely investigated, generalizing the notion of 2 qu-bits. Namely, 
the famous ``Bell theorem without an inequality'' for 3 qu-bits was stated in  \cite{GrHoShZe}, and further discussed in \cite{Me}, 
\cite{Ha}, \cite{Ca}, \cite{HePh};
see \cite{BrCaPiScWe} for a review (up to 2014). 
In \cite{BeFlMa}, \cite{BaCiWo}, entanglement in fermionic systems or quantum metrology is considered from the point of vue of algebra of 
local observables. 
In \cite{KhSpSaDe-Gu} combinatorial properties for multi-mode multi-photonic interferometry are presented.
A comprehensive review on quantum entanglement (up to 2009) is provided in \cite{Ho4}; see also \cite{Ar} for a recent account on quantum cellular 
automata.  

Here we examine in a very elementary way (essentially from the point of vue of tensor algebra) fermionic particles (mostly electrons) 
subject to entanglement 
in passive interferometers, or in polarizer/analyser devices such as a Stern-Gerlach apparatus \cite{Fey}.
In constrast with GHZ or W states, we shall consider only partial entanglement, so that 
the maximum number of intricated fermionic particles with half-integer spin $S$ turns out to exceed by 1 the number of available states,
i.e. equals $2S+2$ instead of $2S+1$. 

For bosons in an interferometer \cite{KhSpSaDe-Gu}, distinguishability plays an important role: while the entangled
particles are distinguishable before entering the device,
the single ones at the output are not. 
Thus we address the problem of encoding, algebraically, the ``collapse'' 
of the wave function of entangled fermionic particles. We limit ourselves with $S=1/2$ (Proposition 2). 
We will not consider here Bell inequalities and quantum correlations for 3 fermions. 

The wave-function of $2S+2$ undistinguishable particles can be represented as a determinant
of order $2S+1$, that we call {\it affine} Slater determinant. We
investigate usual basic properties of $n$-points functions and reduced density matrices.

\section{\bf Intricated particles with spin $S$} 

\smallskip
\noindent {\it 1| Some entangled states: a review}

We discuss according to bosonic or fermionic statistics. 
Recall first the case of photons (or spin-bosons).
For photons the role of spin is played by left/right polarization: $|0\rangle={1\choose0}$, $|1\rangle={0\choose1}$.  
Elementary types of intricated states with 2 qu-bits (bipartite entanglement) $A$ and $B$ are:
\begin{eqnarray*}
&\psi_1^\pm=\frac{1}{\sqrt2} \bigl(|0_A0_B\rangle\pm|1_A1_B\rangle+\perm \bigr)\\ 
&\psi_2^\pm=\frac{1}{\sqrt2} \bigl(|0_A1_B\rangle\pm|1_A0_B\rangle+\perm \bigr)
\end{eqnarray*}
Here $\perm=(A\leftrightarrow B)$ means symmetrized tensor product 
$|0_A1_B\rangle\mapsto \frac{1}{2}\bigl(|0_A1_B\rangle+|0_B1_A\rangle\bigr)$, etc\dots 
We have used the usual notation in tensor calculus: $|0_A0_B\rangle=|0\rangle_A\otimes|0\rangle_B$, in particular 
the two qu-bits are distinguishable: $|0_A0_B\rangle \neq |0_B0_A\rangle$, $|0_A1_B\rangle \neq |0_B1_A\rangle$. 
So $\psi_j^\pm$ are bosonic (invariant under permutation $P$ of A and B). Alternatively, consider 
\begin{equation*}
\phi_2^\pm=\frac{1}{\sqrt2} \bigl(|0_A1_B\rangle\pm|1_A0_B\rangle+|1_B0_A\rangle\pm|0_B1_A\rangle \bigr)
\end{equation*}
So $\phi_2^+$ is again bosonic, but $\phi_2^-$ is fermionic
(changes its sign under permutation $P$ of A and B, where $P$ is the exchange operator). 

Elementary types of intricated states with 3 qu-bits and 3-fold entanglement are 
\begin{eqnarray*}
&|\GHZ\rangle=\frac{1}{\sqrt2} \bigl(|0_A0_B0_C\rangle+|1_A1_B1_C\rangle+\perm\bigr)\\
&|W\rangle=\frac{1}{\sqrt3} \bigl(|0_A1_B1_C\rangle+|1_A1_B0_C\rangle+|1_A0_B1_C\rangle+\perm\bigr)
\end{eqnarray*}
They are invariant under permutation of particles so are again bosonic particles. 
These states are used in Bohm experiments in polarizer/analyser devices, and their quantum correlations violate Bell inequalities.

It is known \cite{Ca} that W states are more ``robust'' than GHZ states,
in the sense that after the 3-fold intrication of a W state is broken there can still
remain a 2-fold entanglement, while this does not hold for GHZ states. 
We may also expect partially entangled states to be more robust than fully entangled states.\\

Consider now bi-partite entanglement for 2 electrons (spin $S=\frac{1}{2}$). 
Fermionic qu-bits are again distinguishable: $|0_A1_B\rangle\neq|0_B1_A\rangle$. For instance we have
\begin{equation*}
\psi_3^\pm=\frac{1}{\sqrt2} \bigl(|0_A1_B\rangle\pm|1_A0_B\rangle+\perm\bigr)
\end{equation*}
where'' perm''  means here anti-symmetrized tensor product 
$|0_A1_B\rangle\mapsto \frac{1}{2}\bigl(|0_A1_B\rangle-|0_B1_A\rangle\bigr)$, etc\dots
So $\psi_3^\pm$ are again fermionic. Alternatively, we can produce $\phi_3^\pm$ as above, such that $\phi_3^+$ is fermionic, 
but $\phi_3^-$ is bosonic. 

From the point of vue of representation of an entangled state however, some identification of tensor products has to be made in order to
account for distinguishability. This will be explained in the next Sect.2 in case of 3 particles.\\

Entanglement of several electrons lead to EPR type experiments with a Stern-Gerlach apparatus.
One of the main purposes is to test maximal violation of Bell inequalities, 
in computing the mean value
of Pauli operators in a given state. For the simplest experiment with 3 particles, the so-called (3,2,2)-scenario,
\cite{Ca1} sets up the table of all 46 possible Bell operators with the corresponding tight Bell inequalities.\\ 

\noindent {\it 2| 2-fold entanglement of 3 fermions of spin 1/2}

Here we are interested in bi-partite entanglement for 3 electrons
(states of the form $ab+bc+ca$, where $ab$ means antisymmetric product, which we shall call a ``3-fermion''), 
or more generally in multi-partite entanglement for
fermionic particles with spin $S$. So for $S=1/2$, this is related with the (3,2,2) scenario of \cite{Ca1}. 
Bell inequalities will be discussed elsewhere.

Consider first the question of distinguishability. To make fermionic qu-bits for 3 particles $A,B,C$ distinguishable.
we need some identifications from the point of vue of tensor algebra. We examine the case $S=1/2$.

Identify copies ${\cal H}_A,{\cal H}_B,{\cal H}_C$ of ${\bf C}^2$ with $E_A={\cal H}_A\oplus0\oplus0$,
$E_B=0\oplus{\cal H}_B\oplus0$, $E_C=0\oplus0\oplus{\cal H}_C$, and the vector $a\in{\cal H}_A$  
with $a'=a\oplus0\oplus0$, etc\dots respectively, 
$a',b',c'\in{\bf C}^6$. In the same way, identify tensor $ab$ with the antisymmetrized product 
$a'\wedge b'=\frac{1}{2}\bigl(a'\otimes b'-b'\otimes a'\bigr)$ ($a'b'$ for short), etc\dots.
Then $a'\wedge b'+b'\wedge c'+ c'\wedge a'\in {\bf C}^{6}\otimes{\bf C}^{6}={\bf C}^{36}$. 
Using coordinates, we write $a={x_A\choose y_A}$, etc\dots. We have by a staightforward computation:\\

\noindent {\bf Lemma 1}: {\it There is a canonical isomorphism 
(reindexation of coordinates) $\theta:{\bf C}^{36}\to{\bf C}^{12}\times{\bf C}^{12}\times{\bf C}^{12}$ such that 
$\Lambda=a'\wedge b'+b'\wedge c'+ c'\wedge a'$ takes the form ($0_p$ denotes the zero vector in ${\bf C}^p$)}
\begin{equation}\label{1}
\theta(\Lambda)=
\bigl({x_A\choose y_A}\otimes \begin{pmatrix}{x_B\choose y_B}\\
{-x_C\choose -y_C}\\ 0_2\end{pmatrix}, 
{x_B\choose y_B}\otimes \begin{pmatrix}{-x_A\choose -y_A}\\ 0_2\\
{x_C\choose y_C}\end{pmatrix}, {x_C\choose y_C}\otimes \begin{pmatrix}0_2\\ {x_A\choose y_A}\\
{-x_B\choose -y_B}\end{pmatrix}\bigr)
\end{equation}

Set $x_{AB}={x_Ax_B\choose x_Ay_B}$, and with similar notations
\begin{equation*}
X'_1=\begin{pmatrix}x_{AB}\\-x_{AC}\\0_2\end{pmatrix}, \ X'_2=\begin{pmatrix}-x_{BA}\\0_2\\x_{BC}\end{pmatrix}, \
X'_3=\begin{pmatrix} 0_2\\x_{CA}\\-x_{CB}\end{pmatrix}
\end{equation*}
In the same way we set $y_{AB}={y_Ax_B\choose y_Ay_B}$ and $(Y'_1,Y'_2,Y'_3)$. 
We write (\ref{1}) as $\theta(\Lambda)=\begin{pmatrix}X'_1&X'_2&X'_3\\  Y'_1&Y'_2&Y'_3\end{pmatrix}$, and 
define the ``partial trace'' 
$$\Tr _1(X'_1,X'_2,X'_3)=X'_1+X'_2+X'_3=(0,x_Ay_B-x_By_A,0,-x_Ay_C+x_Cy_A,0,x_By_c-x_Cy_B)\in{\bf C}^6$$ 
which we identify with the vector of ${\bf C}^3$ consisting in its 3 non-trivial components. 
Since $\Tr _1(Y'_1,Y'_2,Y'_3)=-\Tr _1(X'_1,X'_2,X'_3)$, we skip $(Y'_1,Y'_2,Y'_3)$ henceforth from notations, and define
$\Tr _1\theta(\Lambda)= \Tr _1(X'_1,X'_2,X'_3)$ where $\theta(\Lambda)$ is as in (\ref{1}). 
Consider now the antisymmetric form
$$\omega: {\bf C}^6\to{\bf C}^{36}, \ (a,b,c)\to\Lambda=\wedge(a'\otimes b'+b'\otimes c'+c'\otimes a')$$
and its ``partial trace'' taking values in ${\bf C}^3$
\begin{equation}\label{2}
\omega_1(a,b,c)=\Tr _1\omega(a,b,c)=\Tr _1\theta(\Lambda)
\end{equation}

The ``collapse'' from the distinguishable 3-fermion to the un-distinguishable one is then encoded as follows:\\

\noindent{\bf Proposition 2}: {\it Let $\omega:{\bf C}^6\to{\bf C}^{36}$ 
$$\omega:(a,b,c)\to\Lambda=\wedge(a'\otimes b'+b'\otimes c'+c'\otimes a')$$
and $\omega_1$ 
$$\omega_1(a,b,c)=\Tr _1\omega(a,b,c)=\Tr _1\theta(\Lambda)\in {\bf C}^3$$
where $\Tr _1$ consists in adding the 3 columns of $\theta(\Lambda)$. 
Then there is a projector $\pi:{\bf C}^3\to{\bf C}^3$ of rank 1 such that $\omega_0=\pi\circ\omega_1$ identifies with the antisymmetric 
2-form on ${\bf C}^2$ }
$$\widetilde\omega(a,b,c)=\wedge(a\otimes b+b\otimes c+c\otimes a)=\det (\vec{ab},\vec{ac})$$

\noindent {\it Proof}: 
First notice that if $a=b$, then
$$\omega_1(a,a,c)=(-x_Ay_C+x_Cy_A)\vec u, \quad \vec u=(0,1,-1)$$
and similarly
$$\omega_1(a,b,b)=(x_Ay_B-x_By_A)\vec v, \ \vec v=(1,0,-1), \quad \omega_1(a,b,a)=(x_Ay_B-x_By_A)\vec w,
\vec w=(1,-1,0)$$
In ${\bf C}^3$ consider the 2-plane $\langle \vec u,\vec v\rangle$, the quotient space
${\bf C}^3/\langle \vec u,\vec v\rangle$ is one dimensional.
Let $\pi:{\bf C}^3\to{\bf C}^3/\langle \vec u,\vec v\rangle$ be the canonical projection.  
As $\vec w=\vec v-\vec u\in\langle \vec u,\vec v\rangle$, it readily follows that the form 
$\widetilde\omega=\pi\circ\omega_1$ is well-defined, and identifies with an antisymmetric 2-form valued in ${\bf C}$,
namely the determinant $\det (\vec{ab},\vec{ac})$. $\clubsuit$ 

\smallskip
Thus the form $\widetilde\omega$, which we still denote (abusively) by $\wedge(a,b,c)=\wedge(a\otimes b+b\otimes c+c\otimes a)$
restores the un-distinguishability of particles A,B,C. 
We say that $\widetilde\omega$ is an {\it affine} antisymmetric 2-form, not to be confused with a {\it multilinear} antisymmetric 2-form. 
In the canonical coordinates $(x_A,y_A)$, \dots above we have
\begin{equation}\label{3}
\wedge(a,b,c)=\biggl|\begin{matrix}x_B-x_A&x_C-x_A\\ y_B-y_A&y_C-y_A\end{matrix}\biggr|
\end{equation}

Let $\sigma:{\bf C}^2\to{\bf C}^2$ be a morphism. For the distinguishable particles $a',b',c'$ above, we define 
$(\sigma\otimes\sigma)(a'b'+b'c'+c'a')$ in the representation (1) simply by changing ${x_A\choose y_A}$ to $\sigma{x_A\choose y_A}$, etc\dots
The proof of Proposition 2 readily shows that $(\sigma\otimes\sigma)(a'b'+b'c'+c'a')$ collapses into
$$(\sigma\otimes\sigma)\det(\vec{ab},\vec{ac})=\det(\sigma(\vec{ab}),\sigma(\vec{ac}))=(\det\sigma)\det(\vec{ab},\vec{ac})$$

Notice that $\wedge(a,b,c)=0$ iff the points $a,b,c$ are aligned in ${\bf C}^2$. So if we normalize the vectors $a,b,c$
in ${\cal H}_A={\cal H}_B={\cal H}_C={\bf C}^2$, then $\widetilde\omega(a,b,c)=0$ iff $a=b$ or $a=c$. This is related with
properties of the SU(2) representation of spin 1/2 on Poincar\'e (or Bloch) sphere, which we will not discuss here.\\

\noindent {\it 3| States and partial traces for un-distinguishable 3-fermions}

We define $\rho_{A,B,C}(a,b,c;a',b',c')$ by its matrix elements $\rho_{A,B,C}(a,b,c;a',b',c')=\det(b-a,c-a)\det(b'-a',c'-a')$ in some 
basis of ${\bf C}^2$.
The partial trace of order 1 is given by its matrix elements 
$\Tr _A\rho_{A,B,C}(b,c;b',c')=\sum_a\det(b-a,c-a)\det(b'-a,c'-a)$, where the sum ranges over $a=|0\rangle,|1\rangle$. It is non zero.
Similarly, the partial trace of order 2 is given by
$\Tr _{A,C}\rho_{A,B,C}(b;b')=\sum_{a,c}\det(b-a,c-a)\det(b'-a,c-a)$, and $\Tr _{A,C}\rho_{A,B,C}=0$. 

\section {\bf Affine antisymmetric forms in higher dimensions or of lower degree}

We consider here un-distinguishable fermionic particles (after collapse). So let
${\cal H}={\bf C}^d$,  we recall $\Lambda^p {\cal H}\subset\bigotimes^p {\cal H}$ the space of anti-symmetric,
tensors of degree $p$ of the form
$$\wedge(x_1\otimes\cdots\otimes x_p)=x_1\wedge\cdots\wedge x_p={1\over p!}
\sum_{\sigma\in{\cal S}_p}\eps(\sigma)x_{\sigma(1)}\otimes x_{\sigma(2)}\otimes\cdots\otimes x_{\sigma(p)}$$
which can be identified with antisymmetric $p$-forms.
If $p=d$, $x_1\wedge\cdots\wedge x_d$ is simply the {\it determinant} of $(x_1,\cdots,x_d)$. 
The set $\prod_{p\geq0}\Lambda^p {\cal H}$ is a graded algebra of dimension $2^d$, with graded anti-commutativity
$t\wedge t'=(-1)^{pp'} t'\wedge t$. \\

On ${\cal H}={\bf C}^2$ for instance, 
$a\wedge b=\frac{1}{2}(a\otimes b- b\otimes a)$ is antisymmetric (and bilinear) in $a,b\in {\cal H}$. 
We call it a {\it 2-fermion} (entanglement of $n=2$ particles of spin 
1/2) of degree $p=2$  (degree being understood as degree of homogeneity). 
Instead, the {\it 3-fermion} $a\wedge b+b\wedge c+ c\wedge a$ is antisymmetric (but not bilinear) in $a,b,c$.
We extend this example by introducing\\

\noindent {\bf Definition 3}: The {\it affine determinant} of $x_0,x_1,\cdots,x_d$ in ${\bf C}^d$
is the antisymmetric 2-form in variables $x_0,x_1\cdots,x_d$ 
$$\widetilde\omega(x_0,x_1,\cdots,x_d)=\det(\vec{x_0x_1},\vec{x_0x_2},\cdots,\vec{x_0x_d})$$

A way to find antisymmetric 2-forms is to consider generators. Thus
on ${\cal H}={\bf C}^2$, $\omega(a,b,c)=\wedge(a,b,c)$ is (up to a factor) 
the anti-symmetrized form of $\omega_0(a,b,c)=ab$. We call the monomial $ab$ a {\it generator}. In the same way, 
on ${\cal H}={\bf C}^3$, $\omega_0(a,b,c,d)=abc$ is the generator of the antisymmetrized form 
$$\sum_{\sigma\in{\cal S}_4}\eps (\sigma)\omega(\sigma(a),\sigma(b),\sigma(c),\sigma(d))=6(abc-bcd+cda-dab)$$
which up to a factor, is equal to $\det (\vec{ab},\vec{ac},\vec{ad})$.\\

Let us say that $\omega$ is {\it non-degenerate} on ${\cal H}={\bf C}^d$ iff
$(\forall x_0\in{\cal H} \quad \omega(x_0,x_1,x_2,\cdots,x_d)=0)\Longrightarrow (x_1,\cdots,x_d$ belong to an affine subspace of dimension $d-2$). 
Thus the affine determinant $\omega(a,b,c)=\det(b-a,c-a)$ is, up to a factor, the only non-degenerate form on 
${\bf C}^2$ depending on 3 variables. 
We conjecture that: (1) the affine determinant is
the only non degenerate antisymmetric tensor $\omega$ on ${\bf C}^d$ of degree $d$,
depending on $d+1$ variables, and: (2)
there is no non trivial antisymmetric tensor on ${\bf C}^d$ of degree $d$, depending on more than $d+1$ variables. 

Moreover we conjecture that the affine determinant  results from the collapse (in the sense of Proposition 2) of 
$d+1$ distinguishable fermions of spin $S$, $d=2S+1$ into un-distinguishable fermions.

Instead, making use of Laplace rule 
$$\det(x_1,\cdots,x_n)=\sum_{i=1}^n(-1)^{n+1}x_{i1}{\det }_i(x_2,\cdots,x_n)$$
and its higher order generalization, 
affine determinants extend to all ``partial'' affine determinants of degree $p\leq d-1$.\\

\noindent {\it Example} (symplectic affine form): 
Let $\omega$ the symplectic form on ${\bf R}^{n}$, $n$ even, and
$L_1,L_2,L_3$ Lagrangian subspaces. The 2-form 
$Q=Q(L_1,L_2,L_3)$ on $L_1\oplus L_2\oplus L_3\to{\bf R}$ defined by
$Q(x_1,x_2,x_3)=\omega(x_1,x_2)+\omega(x_2,x_3)+\omega(x_3,x_1)$
is a symplectic affine form (of type $ab+bc+ca$).
Signature of $Q$ is known as Kashiwara index of $(L_1,L_2,L_3)$. 

\section{\bf Affine Slater determinants}

Let $(X, d\mu)$ be a measured space, and 
apply the above discussion to the case where $a_j\in{\bf C}^d$ are functions of $\widehat x=(x_0,x_1,\cdots x_d)\in X^{d+1}$, namely
$a_j=\phi_j(\widehat x)={}^t\bigl(\phi_{1j}(\widehat x),\cdots,\phi_{dj}(\widehat x)\bigr)$, $j=1,2,\cdots,d$. 
Here $\phi_j(\widehat x)\in{\bf C}$ is the $j$:th component of the wave-function,  i.e. $\phi\in L^2(X,d\mu)\otimes{\bf C}^d$. 

In case of affine Slater determinants,
because of ``translation invariance'' (choice of the ``origin'' in the affine space), we need to integrate over one more variable, 
so we shall assume that $d\mu$ is a finite measure, e.g. a probability measure. Physically, we can think of 
of $X$ as a torus (periodic boundary conditions) containing a gas of $d+1$-fermions. \\

\noindent {\bf Definition 4}: The affine Slater determinant of the wave function of $d+1$ particles
$x_0,\cdots,x_d$ is $\Psi(x_0,x_1,\cdots,x_d)=\det\bigl(\varphi(x_1)-\varphi(x_0),\cdots,\varphi(x_d)-\varphi(x_0)\bigr)$, and
$\varphi={}^t\bigl(\varphi_1,\cdots,\varphi_d\bigr)$ has $d$ components. 
\smallskip

\noindent {\it Remark}: it is sometimes convenient to consider the ``centered variable'' 
$\widetilde\varphi(x)=\varphi(x)-\langle\varphi\rangle$ with respect to $d\mu$, $\langle \varphi\rangle=\int_X \varphi(x)\,d\mu(x)$.\\

There follows usual objects such as $n$-point functions, density matrices, and Hamiltonians (see e.g. \cite{Low}, \cite{Ma}). 
For simplicity we restrict
to $S=1/2$, i.e. $d=2$, and assume throughout that $\mu(X)=1$ (in particular if $(X,d\mu)$ is a probability space). Moreover all
wave functions $\varphi$ are supposed to be real valued.\\

\noindent {\it 1| $n$-point functions}\\

The one-point function verifies $\langle\Psi\rangle=0$, and the two-point function
$$\frac{1}{6}\langle\Psi|1|\Psi\rangle=\bigg|
\begin{matrix}\langle\widetilde\varphi_1^2\rangle&\langle\widetilde\varphi_1\widetilde\varphi_2\rangle\\
\langle\widetilde\varphi_1\widetilde\varphi_2\rangle&\langle\widetilde\varphi_2^2\rangle\end{matrix}
\bigg|$$
which equals 1 if $\varphi_j$ are normalized and orthogonal in the Hilbert space $L^2(X,d\mu)$. 
If $M(x_0,x_1,\cdots,x_d)$ is symmetric in its arguments, we have also (with obvious notations)
$$3\int ab M(x)(ab+bc+ca)\,d\mu(x)=\int (ab+bc+ca)M(x)(ab+bc+ca)\,d\mu(x)$$
In higher dimensions, such formula generalize to overlaps of 
affine Slater determinants. \\

\smallskip
\noindent {\it 2| Density matrices}\\

We can generalize classical results [Low] to affine Slater determinants. For simplicity, assume $d=2$. 
We define the density matrix of order 1
\begin{eqnarray*}
\Gamma(x'_1,x_1)&=\int \int \bigg|\begin{matrix}\widetilde\varphi_1(x_1)-\widetilde\varphi_1(x_0)&\widetilde\varphi_1(x_2)-\widetilde\varphi_1(x_0)\\
\widetilde\varphi_2(x_1)-\widetilde\varphi_2(x_0)&\widetilde\varphi_2(x_2)-\widetilde\varphi_2(x_0)
\end{matrix}\bigg|\times\\
&\bigg|\begin{matrix}\widetilde\varphi_1(x'_1)-\widetilde\varphi_1(x_0)&\widetilde\varphi_1(x_2)-\widetilde\varphi_1(x_0)\\
\widetilde\varphi_2(x'_1)-\widetilde\varphi_2(x_0)&\widetilde\varphi_2(x_2)-\widetilde\varphi_2(x_0)\end{matrix}\bigg|\,d\mu(x_0)\,d\mu(x_2)
\end{eqnarray*}
which we normalize by
$$\gamma^{(1)}(x'_1,x_1)={1\over\mu(X)}\biggl({1\over2}\Gamma(x'_1,x_1)-\bigg|
\begin{matrix}\langle\widetilde\varphi_1^2\rangle&\langle\widetilde\varphi_1\widetilde\varphi_2\rangle\\
\langle\widetilde\varphi_1\widetilde\varphi_2\rangle&\langle\widetilde\varphi_2^2\rangle\end{matrix}\bigg|\biggr)$$
(Hermitian matrix). In centered reduced variables $\widetilde\varphi_j$
$\gamma^{(1)}(x'_1,x_1)=\widetilde\varphi_1(x'_1)\widetilde\varphi_1(x_1)$, which is the 
analogue of density matrix of order 1 for the usual Slater determinant. \\

We define the density matrix of order 2, by integrating only over $x_0$:
\begin{equation*}
\gamma^{(2)}(x'_1,x'_2;x_1,x_2)=
\int \bigg|\begin{matrix}
\widetilde\varphi_1(x_1)-\widetilde\varphi_1(x_0)&\widetilde\varphi_1(x_2)-\widetilde\varphi_1(x_0)\\
\widetilde\varphi_2(x_1)-\widetilde\varphi_2(x_0)&\widetilde\varphi_2(x_2)-\widetilde\varphi_2(x_0)\end{matrix}\bigg|
\bigg|\begin{matrix}\widetilde\varphi_1(x'_1)-\widetilde\varphi_1(x_0)&\widetilde\varphi_1(x'_2)-\widetilde\varphi_1(x_0)\\
\widetilde\varphi_2(x'_1)-\widetilde\varphi_2(x_0)&\widetilde\varphi_2(x'_2)-\widetilde\varphi_2(x_0)\end{matrix}\bigg|\,d\mu(x_0)
\end{equation*} 
This density matrix is Hermitian, and anti-symmetric in each set of indices
\begin{eqnarray*}
&\gamma^{(2)}(x'_1,x'_2;x_1,x_2)=\gamma^{(2)}(x_1,x_2;x'_1,x'_2)\\
&\gamma^{(2)}(x'_1,x'_2;x_1,x_2)=-\gamma^{(2)}(x'_2,x'_1;x_1,x_2)
\end{eqnarray*}
For reduced centered variables, this is a positive definite matrix, since
\begin{eqnarray*}
&\gamma^{(2)}(x'_1,x'_2;x_1,x_2)=\bigl(\widetilde\varphi_2(x_1)-\widetilde\varphi_2(x_2)\bigr)
\bigl(\widetilde\varphi_2(x'_1)-\widetilde\varphi_2(x'_2)\bigr)+\\
&\bigl(\widetilde\varphi_1(x_1)-\widetilde\varphi_1(x_2)\bigr)\bigl(\widetilde\varphi_1(x'_1)-\widetilde\varphi_1(x'_2)\bigr)+
\bigg|\begin{matrix}\widetilde\varphi_1(x_1)&\widetilde\varphi_1(x_2)\\ \widetilde\varphi_2(x_1)&\widetilde\varphi_2(x_2)\end{matrix}\biggr|
\bigg|\begin{matrix}\widetilde\varphi_1(x'_1)&\widetilde\varphi_1(x'_2)\\ \widetilde\varphi_2(x'_1)&\widetilde\varphi_2(x'_2)\end{matrix}\bigg|
\end{eqnarray*}
The last term is the usual one, and first two ones remind of the affine structure.
As in \cite{Low}, we can expand a $n$-body Hamiltonian as 
$$\Omega=\Omega_0+\sum_i\Omega_i+\frac{1}{2!}\Omega_{i\neq j}+\frac{1}{3!}\Omega_{i\neq j\neq k\neq i}\Omega_{ijk}+\cdots$$
and compute the mean value of $\Omega$ in the state $\Psi$, using density matrices. This applies in particular to 
the total spin $S^2$ for antisymmetric particles 
$\Omega=S^2=\sum_{i,j=1}^3 \sigma_i\cdot \sigma_j$ (Pauli matrices), which we express as the exchange Hamiltonian
$P={1\over2}(\Id+\sigma_1\cdot\sigma_2)$, $P|ij\rangle=|ji|\rangle$. \\

\noindent{\bf Acknowledgements}: Special thanks are due to Ad\`an Cabello for useful informations.\\

\end{document}